# Design of a High Speed FPGA-Based Classifier for Efficient Packet Classification

V.S.Pallavi[1], Dr.D.Rukmani Devi[2]

PG Scholar[1], Department of ECE, RMK Engineering College, Chennai, Tamil Nadu, India
Professor[2], Department of ECE, RMK Engineering College, Chennai, Tamil Nadu, India

*Abstract*— **Packet classification is a vital and complicated task as the processing of packets should be done at a specified line speed. In order to classify a packet as belonging to a particular flow or set of flows, network nodes must perform a search over a set of filters using multiple fields of the packet as the search key. Hence the matching of packets should be much faster and simpler for quick processing and classification. A hardware accelerator or a classifier has been proposed here using a modified version of the HyperCuts packet classification algorithm. A new pre-cutting process has been implemented to reduce the memory size to fit in an FPGA. This classifier can classify packets with high speed and with a power consumption factor of less than 3W. This methodology removes the need for floating point division to be performed by replacing the region compaction scheme of HyperCuts by pre-cutting, while classifying the packets and concentrates on classifying the packets at the core of the network.**

*Keywords*— **Classifier, high throughput, low power, packet classification, pipelining, region compaction, pre-cutting.**

## I. INTRODUCTION

The necessity of packet classification is considered important as the burden of a router is reduced. Initially the task of putting a real strain on the networking equipment has to be inspected and processed to resist the resultant traffic. Packet classification continues to grow in importance, both at the edge and the core. Existing algorithms still have poor performance, and ternary CAMs still have issues in terms of power consumption and chip density [1]. Despite the large number of ideas explored, there are still new ideas in packet classification that can provide major benefits.

Packet classification algorithms use two dominant resources, memory and time. Reducing worst-case search time in memory references is equally important. Network processors are key components used to process packets as they pass through a network, carrying out tasks such as packet fragmentation and reassembly, encryption, forwarding, and classification. The increase in line rates, have placed the network processor under increased pressure and the inevitable role of the classifier here is to reduce the above mentioned heavy tasks of it without any major degradation in speed, power and area. The hardware accelerator can be designed to have fewer transistors than that of the general-purpose processors used in multi-core network processors.

Hardware accelerators can also process more data than a general-purpose processor while running at slower clock speeds as they are optimized to carry out specific tasks. A reduction in clock speed and number of transistors leads to large savings in power consumption and area. The rest of the paper is organized as follows. Section II explains the pre-cutting based packet classification and gives a detailed explanation of the HyperCuts algorithm. This is done so that the changes made here to make the algorithm more suited to hardware acceleration can be better understood. Section III explains the architecture of the classifier. The performance results including the memory usage, throughput and power consumption are given in section IV. Section V concludes the paper.

## II. PRE-CUTTING BASED PACKET CLASSIFICATION

The fields of a packet's header most commonly used to perform packet classification are the 32 b source and destination IP addresses, the 16 b source and destination port numbers, and the 8 b protocol number.

The information presented in this paper centers around the design and implementation of an energy-efficient packet classification hardware accelerator (classifier) that can relieve a network processor's processing engines of the difficult and power hungry networking task of packet classification.

Pre-cutting scheme only requires an internal or root node to store the number of cuts that must be performed to each field of a packet header and the bits in these fields where the cuts are to be performed [3]. The simplicity of this scheme helps to improve throughput and decrease power consumption. The region that needs to be divided is compacted by recursively cutting all fields in two. Each precut to a field used to divide the region will halve the number of sub-regions that need to be stored and the number of cuts that need to be performed to a packet header when selecting the sub-region to traverse. Cutting of an internal node can be done more efficiently by using these steps,

**Step A:** Perform precuts to the source and destination IP addresses as shown in figure below, which reduces the area for cutting by 75%. Precuts can be performed to both fields.

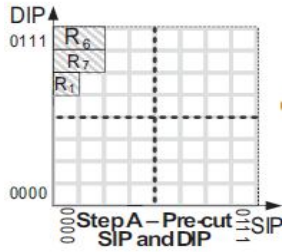

**Step B:** Only the source IP is precut as pre-cutting the destination IP would result in more than one sub-region that contains rules as shown in figure below.
Pre-cutting the source IP reduces the area by another 50 %.

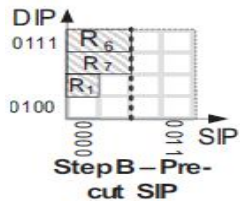

**Step C:** In step C no more precuts can be performed so the compacted region is cut in two, with none of the resulting sub-regions containing more than two rules as shown in figure below.

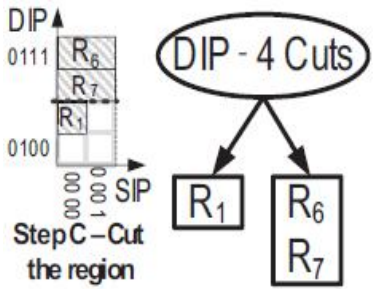

The classifier presented by this cutting scheme allows packet classification to be moved to the core of a network, thus improving security [14]. It uses multiple packet classification engines working in parallel with a shared memory, allowing it to classify packets at speeds of several Gb/s, while using rulesets containing tens of thousands of rules. It implements a modified version of the HyperCuts packet classification algorithm, which breaks a ruleset into groups, with each group containing a small number of rules that can be searched linearly.

## II. HEURISTICS IN MODIFIED HYPERCUT ALGORITHM

The HyperCuts packet classification algorithm uses different heuristics to reduce the amount of memory needed to save a decision tree and the number of memory accesses required to match a rule. This section gives a brief description of these heuristics.

### A. Memory Usage

One of these heuristics is called node merging and it is used to avoid the duplicated storage of identical nodes. Node merging is carried out by first searching the decision tree for leaf nodes that contain the same list of rules. The pointers to these nodes (stored in root and internal nodes) are then modified so that they point to just one of these leaf nodes, meaning that multiple copies do not need to be stored [11]. A second heuristic called rule overlap is used to avoid the storage of rules in leaf nodes that can never be matched. A rule can never be matched and is, therefore, removed from a leaf node if the hypercube of a rule with a higher priority completely covers the space it occupies within the leaf node's subregion. A third heuristic used to avoid the duplicated storage of rules is called pushing common rule subset upward. This heuristic stores rules at an internal or root node that would otherwise need to be stored in all of the internal or root node's subregions.

### B. Rule Storage

Modifications have also been made to the way the rule is stored in a leaf node to reduce both memory consumption and the number of memory accesses needed to retrieve the information required to match a packet header to a rule.
1) The first modification is to store the actual rule in the leaf node rather than a pointer to the rule. This was found during testing of rulesets to have only a small increase in memory consumption for some rulesets and a reduction for others as pointers to rules do not need to be stored. Storing the actual rule rather than a pointer to it allows for a large increase in throughput as data are presented to the classifier one clock cycle earlier.
2) A second modification is to reduce the number of bits required to store the source and destination IP addresses from 76 b down to 70 by using an encoding scheme. An IP address usually requires 32 b to store its address and 6 b to store its mask [12]. The mask number is used to specify the number of MSBs of the address that must be an exact match to the corresponding bits in a packet header to record a match. The remaining LSBs are wildcard bits, meaning that the value of the corresponding bits in a packet header can have any value and still record a match. The encoding scheme stores the 32 b IP address and 6 b masks as a 35 b number.

The least significant bit is used to indicate if more than 28 b of the IP address need to be matched exactly. If not set, 32 are used to store the IP address, with the remaining 2b indicating the number of bits that need to be matched.

## III. ARCHITECTURE OF THE CLASSIFICATION ENGINE

The block diagram of our packet classification hardware accelerator consists of two major blocks namely Tree Traversing and leaf node searcher. When the five fields of an incoming packet have been extracted, they are used by the Tree Traverser to find the node corresponding to the subspace represented by their values.

   If the final child node is empty, the packet classification engine reports Unmatch and gets ready for the processing of next incoming packet [9]. Otherwise, the address information of the leaf node is sent to the Leaf Node Searcher. In our implementation, the rules are compared with the five packet fields by hardware logic in parallel to speed up the searching process. If the rules spread in multiple words, the Leaf Node Searcher continues the comparisons until either a matching rule is found or the last rule in the leaf node is encountered.
In this implementation, we mainly target FPGA devices, with their internal block memory being used for decision tree data structure storage.

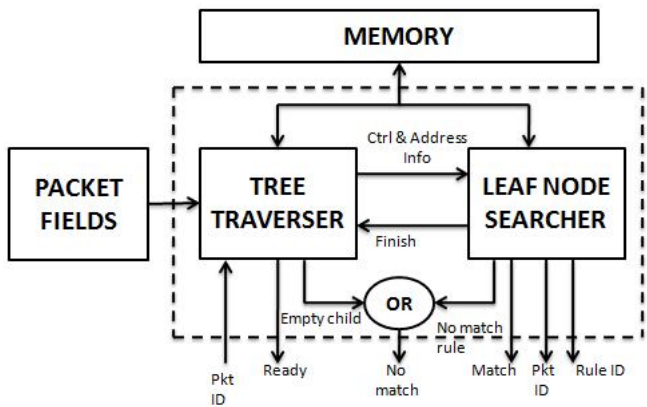

Fig.1.   Architecture of the Packet Classification Engine

One advantage that FPGA devices can provide is the flexible memory word length, which does not have the constraints of external memory chips such as pin numbers and fixed data width. Almost every step of the packet classification procedure involves memory access, either tree structure or the rules in leaf node [7]. This makes the design quite memory-centric, i.e. our major consideration is to reduce the number of memory accesses, without sacrificing too much memory utilization.
On the other hand, ultra-wide memory word reduces the maximum possible working frequency of the packet classification engine. To increase the throughput, we also need to make a balance between the lengths of memory word and the clock rate. The implementation described in this paper is fetched in one memory access with an acceptable working frequency. The functioning of the two major blocks such as Tree Traverser and Leaf Node Searcher is explained below.

### A. Tree Traverser

The first module is a tree traverser that is used to traverse a decision tree using header information from the packet being classified. The decision tree is traversed until an empty node is reached, meaning that there is no matching rule, or a leaf node is reached. Since each incoming packet goes through the root node, the root node information is stored in registers of the packet classification engine, to save one memory access for each classification. Information on the decision tree's root node is stored in registers in the tree traverser, making it possible for the tree traverser to begin classifying a new packet while the previous packet is being compared with rules in a leaf node. This use of pipelining allows for a maximum throughput of one packet every two clock cycles if the decision tree is made up of only a root node and leaf nodes containing no more than two rules.

### B. Leaf Node Searcher

A leaf node being reached will result in the tree traverser passing the packet header and information about the leaf node reached to the second module known as the leaf node searcher. The leaf node searcher compares the packet header to the rules contained in the leaf node until either a matching rule is found or the end of the leaf node is reached with no rule matched. The leaf node searcher employs two comparator blocks that work in parallel [8]. This allows two rules to be searched on each memory rest 6 bits indicating the actual length. If not, all the 32 bits are used for IP with the rest two bits encoded to represent the four possible lengths.

## IV. PACKET CLASSIFICATION HARDWARE ACCELERATOR

A diagram of the hardware accelerator using 4 classification engines working in parallel is shown in figure 2. Then the block RAM runs at a speed that equals to the sum of each engine. The frequency of each engine has a different phase in order to make sure that on every clock cycle of the block RAM, it receives an access requirement. Each packet classification engine will assert a *match* or *no match* signal every time it has finished classifying a packet. This *match* and *no match* signal along with the corresponding matching rule number and packet tag from each of the packet classification engines are multiplexed together. They are then inputted into a logic block used to sort out the matching rule numbers from the engines, so that they are outputted from the hardware accelerator in the correct sequences. The sorter logic block consists of a chain of 16 registers and 15 multiplexers in series. Control logic will register a matching rule number or blank rule number to register when a *match* or *nomatch* access, reducing lookup times. Each rule requires 1 bit for determining if it is the last rule in a leaf node. If no rule has been matched by the time this flag is met set, the Leaf Node Searcher reports *No Matching Rules* and stops working for the current packet. Each rule also requires a 16-bit *Rule ID*. For the protocol field, 8 bits are used to store the protocol number and 1 bit for the mask [2]. Both the source and destination ports require 16 bits for each of two boundaries of their ranges. The source and destination IP addresses use 35 bits to represent the prefixes. The lowest bit is used to indicate whether the prefix length is smaller than 28. If so, only 28 bits are needed to store the IP address, with signal is asserted. The register selected will depend on the packet tag number. The rule number will be registered to the output register if it is next in the sequence of packet results to be outputted and stored if not. All stored rules will be shifted towards the output register each time a rule appears which is due to be outputted. This process is hidden, with the hardware accelerator outputting the result of classified packets on a first come, first served basis.

The classifier has been implemented with multiple packet classification engines working in parallel [4]. The maximum clock speed that an engine can achieve when implemented using an FPGA is much slower than the maximum clock speed of a FPGAs internal memory. This is due to logic delays in the components used by an engine such as the comparator blocks. It is, therefore, necessary to use multiple engines working in parallel so that the classifier can achieve maximum throughput. The use of multiple engines will help to ensure that the bandwidth of a FPGAs internal memory is better utilized. Then the RAM runs at a speed that equals to the sum of each engine. The frequency of each engine has a different phase in order to make sure that on every clock cycle of the block RAM, it receives an access requirement.

Each classifier reads data from a separate data port and has its own packet buffer for storing the headers of incoming packets, four engines that work in parallel to maximize the bandwidth usage of a data port and a sorter logic block used to make sure that the classification results are outputted in the correct order [13]. The packet buffer stores the source and destination IP addresses, source and destination port numbers, and protocol number from the incoming packets. It works on a first come, first served basis, with packets being outputted from the buffer to the packet classification engines in the same order that they were inputted. The buffer also creates a packet ID for each header that is passed to the packet classification engine along with the packet header. The packet ID is used to make sure that the matching rule IDs are outputted by the classifier in the same order that the packet headers were inputted to the system. On the next rising clock edge the hardware accelerator checks if a matching rule has been found. The hardware accelerator will continue searching the leaf node if a matching rule has not been found.

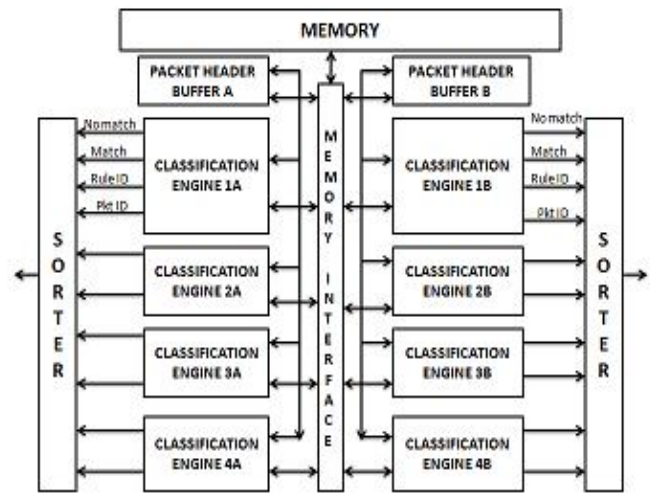

Fig. 2   Architecture of the hardware accelerator

The sorter logic block and multiplexers used for multiplexing the output signals of the classification engines are not needed.

## V. Performance Analysis

The classifier can be tested extensively by measuring its logic and memory usage, throughput in terms of Mpps (millions of packets per second), amount of memory it requires when storing the search structures needed to classify packets. The simulation results of the classifier are brought out using ModelSim as it is an UNIX, Linux and Windows-based simulation-debug environment, combining high performance with the most powerful and intuitive GUI in the industry.

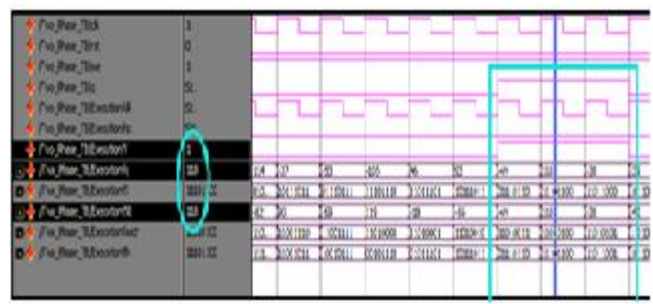

Fig. 3   Packet Drop condition of the hardware accelerator

The function of a classifier is to classify packets according to a specific criterion and either forward those classified packets or drop them.   In accordance to the above condition, the waveform results are shown above in figure 3. The packet drop condition of a classifier can be explained if there is a spike or a high level transition on the signal 'V' of the output. This condition prevails only if the 'q' value is equal to the 'search_exact' value, leading to the drop in the packet that has to be transmitted or classified. The waveform showing only the selected portions of the packet drop condition is shown. The values showing the presence of a spike is encircled at the left and the corresponding waveforms at the right. The Architecture was synthesized using Altera's Quartus 2 design software along with the memory and the logic resources that are needed to implement the hardware accelerator.

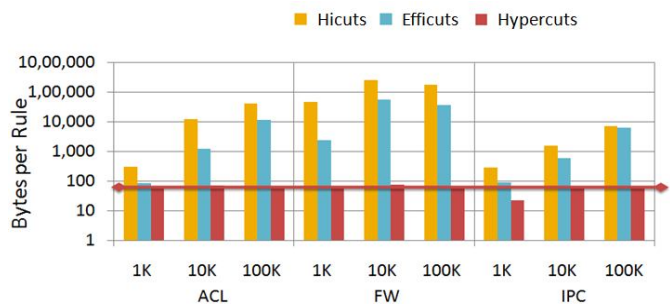

Fig. 4   Graph showing algorithmic efficiency

The search structures built using the ACL, IPC rulesets for both the Cyclone and Stratix implementation show similar performance results. This is because memory consumption is not a major problem for these rulesets as they don't contain many rules with wildcard fields.

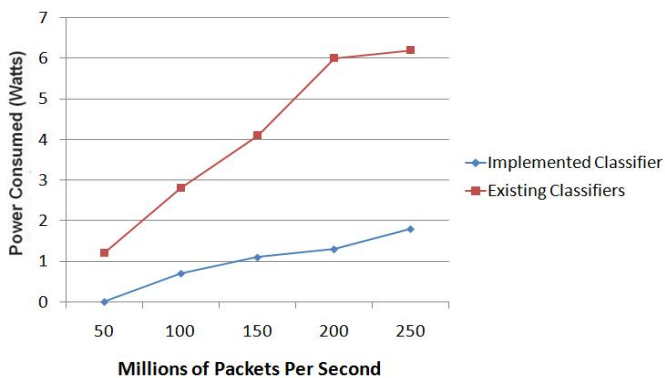

Fig. 5   Graph showing Power Consumed by the classifier

For the search structure built using the FW rulesets it can be seen that the Cyclone implementation shows better performance than the previous FPGA implementations. This is because memory consumption is a problem due to the many rules containing wildcard fields. The power consumed by the hardware accelerator is shown in the graph (figure 4). Post place and route simulations were carried out using Quartus 2 PowerPlay Power Analyzer Tool with VCD files generated by

ModelSim. These results were compared to the power consumed by the state of the art Cypress Ayama 10000 Network Search Engine, which uses similar amounts of memory. The hardware accelerator implemented on the Cyclone 3 FPGA is better than the previously implemented technologies [15]. As modified Hypercuts algorithm improves upon Hicuts and Efficuts algorithmic techniques in terms of both memory and throughput, it is the most effective decision tree algorithm. As packet classification relies on finding the highest priority rule that matches a network packet, it is the key for many vital areas such as security, QOS, traffic monitoring and analysis.

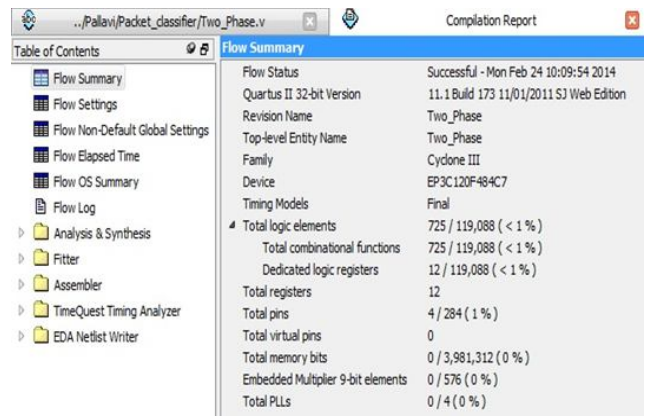

Fig. 6   Minimal usage of logic elements through Cyclone III

A comparison was performed on all the families of Cyclone, by choosing the third family to suit the needs of this classification which is shown in figure 6. Lesser usage of logic elements leads to a decrease in the area consumed, thereby leading to the satisfaction of the major criteria in VLSI. On account of this, the throughput obtained was found to be greater than 220 Mpps.

## VI. CONCLUSION

Packet classification was implemented using a hardware accelerator with enough processing power to allow packet classification to be implemented at the core of a network, thus improving security. The classifier consumed only about 3 W when classifying packets at its maximum throughput of above 220 Mpps by using a modified version of the HyperCuts algorithm so that it is better suited to hardware implementation. This is low when compared to other FPGA-based classifiers. These modifications included changing the cutting scheme so that the need for slow and logic intensive floating point division is removed when classifying a packet. This was done by replacing the region compaction scheme used by HyperCuts with a new scheme that uses pre-cutting.